\documentclass[structabstract]{aa}
\usepackage{graphicx}
\usepackage{txfonts}
\usepackage{longtable}

\begin{document}
   \title{Detection of the 128 day radial velocity variations in the supergiant $\alpha$ Persei}
   \subtitle{Rotational modulations, pulsations, or a planet?}
   \author{Byeong-Cheol Lee\inst{1},
          Inwoo Han\inst{1},
          Myeong-Gu Park\inst{2},
          Kang-Min Kim \inst{1},
          \and
          David E. Mkrtichian\inst{3}
          }

   \institute{Korea Astronomy and Space Science Institute, 776,
		Daedeokdae-Ro, Youseong-Gu, Daejeon 305-348, Korea\\
	      \email{[bclee;iwhan;kmkim]@kasi.re.kr}
	    \and
	      Department of Astronomy and Atmospheric Sciences,
	      Kyungpook National University, Daegu 702-701, Korea\\
	      \email{mgp@knu.ac.kr}
        \and
             Crimean Astrophysical Observatory,
             Nauchny, Crimea, 98409, Ukraine\\
             \email{davidm@crao.crimea.ua}
             }

   \date{Received 29 November 2011 / Accepted 11 May 2012}


  \abstract
   {}
   {In order to search for and study the nature of the low-amplitude and long-periodic radial velocity (RV) variations of massive stars, we have been carrying out a precise RV survey for supergiants that lie near or inside the Cepheid instability strip.
   }
   {We have obtained high-resolution spectra of $\alpha$ Per (F5 Ib) from November 2005 to September 2011 using the fiber-fed Bohyunsan Observatory Echelle Spectrograph (BOES) at Bohyunsan Optical Astronomy Observatory (BOAO).
   }
   {Our measurements reveal that $\alpha$ Per shows a periodic RV variation of 128 days and a semi-amplitude of 70 m s$^{-1}$. We find no strong correlation between RV variations and bisector velocity span (BVS), but the 128-d peak is indeed present in the BVS variations among several other significant peaks in periodogram.
   }
   {
   $\alpha$ Per may have an exoplanet, but the combined data spanning over 20 years seem to suggest that the 128-d RV variations have not been stable on long-term scale, which is somewhat difficult to reconcile with the exoplanet explanation. We do not exclude the pulsational nature of the 128-d variations in $\alpha$ Per. Although we do not find clear evidence for surface activity or rotational modulations by spots, coupled with the fact that the expected rotation period is $\sim$ 130 days the rotational modulation seems to be the most likely cause of the RV variations. More observational data and research are needed to clearly determine the origin of RV the variations in $\alpha$ Per.
   }


   \authorrunning{Byeong-Cheol Lee et al.}
   \titlerunning{Detection of the 128 day radial velocity variations in the supergiant $\alpha$ Persei}
   \maketitle
%

\section{Introduction}

During their evolution, massive stars undergo pulsations by the $\kappa$-mechanism, which operates in the helium ionization zone. At this stage, their expanding outer layers become unstable and they pulsate regularly
within a certain region in the H-R diagram, known as the Cepheid instability strip.
Depending on the mass, stars can move across the instability strip and back again several times as they continue to evolve. Classical Cepheids (Cepheids I) and W Virginis stars (Cepheids II) lie at the intersection of
the strip with the supergiant branch in metal-rich and metal-poor stars.

Several studies have shown that at least one-half of the stars that reside near the Cepheid instability strip are photometrically $\it stable$ at the level of 0.01 $\sim$ 0.03 magnitude (Fernie \& Hube 1971; Percy 1975; Fernie 1976; Percy et al. 1979; Percy \& Welch 1981). However, among the $\it stable$ non-variable stars, low-amplitude variations in the radial velocity (RV) have been discovered thanks to high-resolution spectroscopy (Butler 1992; Hatzes \& Cochran 1995). Despite these efforts, the nature of the low-amplitude variations and the reason for the existence of non-variable stars in the Cepheid instability strip are not yet known. Low-amplitude variations outside the blue edge of the instability strip may be common, but until now this has been scarcely investigated. Therefore, finding RV variations in Cepheids-like stars near the strip may help to understand these stars.

The bright F supergiant $\alpha$ Per (33 Persei, HD 20902, HR 1017, SAO 38787, HIP 15863) in the well-studied cluster of the same name has been observed for photometric and spectroscopic variations since the end of the 1890s.
A history of RV determinations of $\alpha$ Per recorded before the development of precise RV technique is listed in Table~\ref{tab1}. There are few observed RV results of $\alpha$ Per despite the long interval. These show that the RV values vary with a range of $\sim$ 2 km s$^{-1}$, which indicates the possibility of periodic RV variations.

In this paper, we present new precise RV measurements of $\alpha$ Per acquired in 2005 -- 2011. In Section 2, we describe the properties of $\alpha$ Per. We analyze the RVs and search for variability in Section 3. The origin of the RV variations is discussed in Section 4, and we discuss our results in Section 5.

%
\begin{table}
\begin{center}
\caption{The RV measurements of $\alpha$ Per recorded before the development of high-precision RV measurements.}
\label{tab1}
\begin{tabular}{cc}
\hline
\hline
    RV                   &  References      \\
    (km s$^{-1}$)        &                  \\
\hline
    --2.40 $\pm$ 0.53    & Campbell (1898)  \\
    --2.12 $\pm$ 0.4             & Campbell (1901)    \\
    --3.22 $\pm$ 0.69    & Vogel (1901)    \\
    --2.69 $\pm$ 0.79    & Belopolsky (1904)  \\
    --2.24               & K{\'u}stner (1908)    \\
    --1.66 $\pm$ 0.61    & Goos (1909)    \\
    --1.57 $\pm$ 1.43    & Belopolsky (1911)    \\
    --3.42 $\pm$ 0.87    & Hnatek (1912)  \\
    --3.0 $\pm$ 1.0      & Wilson \& Joy (1952)  \\
    --2.4 $\pm$ 0.9      & Wilson  (1953)  \\
    --2.8 $\pm$ 0.5      &  Parsons (1983)  \\
    --2.9 $\pm$ 0.4      &  Beavers \& Eitter (1986)  \\
    --2.08 $\pm$ 1.62    &  Mkrtichian (1990) \\
\hline
\end{tabular}
\end{center}
\end{table}
%

%

\section{The properties of the F supergiant $\alpha$ Per}

%
\begin{table}
\begin{center}
\caption[]{Stellar parameters of $\alpha$ Per.}
\label{tab2}
\begin{tabular}{lcc}
\hline
\hline
    Parameter          & Value      &    Reference     \\

\hline
    Spectral type            & F5 Ib     & Hipparcos  \\
    $\textit{$m_{v}$}$ [mag]  & 1.82     & Hipparcos  \\
    $\textit{$M_{v}$}$ [mag]  & -- 4.67  & Hipparcos \\
    $\textit{B-V}$ [mag]     & 0.49     & Mermilliod (1986) \\
    age [Myr]                &   41       &  Lyubimkov et al. (2010)  \\
    Distance [pc]            & 156 $\pm$ 4    & Lyubimkov et al. (2010) \\
    RV [km s$^{-1}$]         & -- 2.0  $\pm$ 0.1 & Gontcharov (2006) \\
                             & -- 3.0  $\pm$ 1.5 & Kudryavtsev et al. (2007) \\
    Parallax [mas]           & 6.44 $\pm$ 0.17  & van Leeuwen (2007) \\
                             & 6.43 $\pm$ 0.17  & Lyubimkov et al. (2010) \\
    Diameter [mas]           & 3.20 $\pm$ 0.25 & Koechlin \&  Rabbia (1985) \\
                             & 2.8\tablefootmark{a}, 2.9\tablefootmark{b} & Pasinetti-Fracassini et al. (2001) \\
    $T_{\mathrm{eff}}$ [K]   & 6270 $\pm$ 120      & Evans et al. (1996) \\
                             & 6240 $\pm$ 20      & Lee et al. (2006) \\
                             & 6541 $\pm$ 84      & Kovtyukh (2007) \\
                             & 6350 $\pm$ 100      & Lyubimkov et al. (2010) \\
    $\mathrm{[Fe/H]}$       & -- 0.28 $\pm$ 0.06 & Lee et al. (2006) \\
                            & -- 0.07 $\pm$ 0.09 & Lyubimkov et al. (2010) \\
    log $\it g$             & 1.5                   & Evans et al. (1996) \\
                            & 0.58  $\pm$ 0.04      & Lee et al. (2006) \\
                            & 2.0                   & Kovtyukh et al. (2008) \\
                            & 1.90  $\pm$ 0.04      & Lyubimkov et al. (2010) \\

    $\textit{$R_{\star}$}$ [$R_{\odot}$] & 60.7          & Spaan et al. (1987) \\
                                         & 55\tablefootmark{a}, 51\tablefootmark{b} & Pasinetti-Fracassini et al. (2001) \\
    $\textit{$M_{\star}$}$ [$M_{\odot}$] & 7.2           & Parsons \& Bouw (1971) \\
                                         & 7.2           & Becker et al. (1977)  \\
                                         & 8.4           & Spaan et al. (1987)  \\
                                         & 7.3 $\pm$ 0.3 & Lyubimkov et al. (2010)  \\
    $\textit{$L_{\star}$}$ [$L_{\odot}$] & 5500            & Spaan et al. (1987)   \\
    $v_\mathrm{rot}$ sin $i$ [km s$^{-1}$] &  17.9 $\pm$ 1.0    & Gray \& Toner (1987) \\
                                    &  18              & Hatzes \& Cochran (1995) \\
                                    &  17.1 $\pm$ 1.0  &  de Medeiros et al. (2002) \\
                                    &  20              & Lee et al. (2006) \\
    $P_\mathrm{rot}$ / sin $i$ [days]      &  129 -- 139   &  Derived \\
    $v_\mathrm{micro}$ [km s$^{-1}$]       & 3.20 $\pm$ 0.05 & Lee et al. (2006) \\
                                    & 5.3 $\pm$ 0.5 & Lyubimkov et al. (2010) \\
    $v_\mathrm{macro}$ [km s$^{-1}$]       & 10            & Hatzes \& Cochran (1995) \\

\hline

\end{tabular}
\end{center}
\tablefoottext{a}{Intensity interferometer}
\tablefoottext{b}{Intrinsic brightness and color}
\end{table}

The improved investigation of the physical parameters of $\alpha$ Per was reported by Evans et al. (1996).
They measured $\alpha$ Per by comparing its energy distribution derived from IUE spectra and B, V, R, I, J, and K with model atmospheres convolved with instrumental sensitivity functions.
The main parameters were reported by Lyubimkov et al. (2010), who used spectroscopic and the photometric methods and improved Hipparcos parallaxes by van Leeuwen (2007) to determine the fundamental parameters.
The compiled basic stellar parameters of $\alpha$ Per are summarized in Table~\ref{tab2}.

Stellar radius and diameter are taken from the result of Pasinetti-Fracassini et al. (2001).
They calculated them with several methods using an intensity interferometer (Hanbury Brown et al. 1967)
and intrinsic brightness and color (Wesselink 1969; Blackwell \& Shallis 1977).
Because the stellar mass has been determined somewhat differently by individual authors,
the final mass of a companion (for exoplanet explanation) may vary.
Table~\ref{tab2} shows the similar determinations of stellar mass even though Spaan et al. (1987)
estimated a somewhat larger mass of 8.4 $M_{\odot}$.

It is difficult to determine a true rotational period because of an inaccurate stellar radius and ignorance of the stellar inclination. Nonetheless, the values of the projected rotational velocity, $v_\mathrm{rot}$ sin $i$, in the literature are consistent. Based on the recently published value of 20 km s$^{-1}$ (Lee et al. 2006) and a stellar radius of 51 -- 55 $R_{\odot}$ (Pasinetti-Fracassini et al. 2001),
we derived a range of the upper limit of the rotational period of

   \begin{equation}
      P_{\mathrm{rot}} = 2 \pi R_{\star} / (v_\mathrm{rot} \sin \emph{i})  = 129 - 139 \,\mathrm{days} \,.
   \end{equation}

%

\section{Observations and analysis}

%
   \begin{figure*}
   \centering
   \includegraphics[width=17cm]{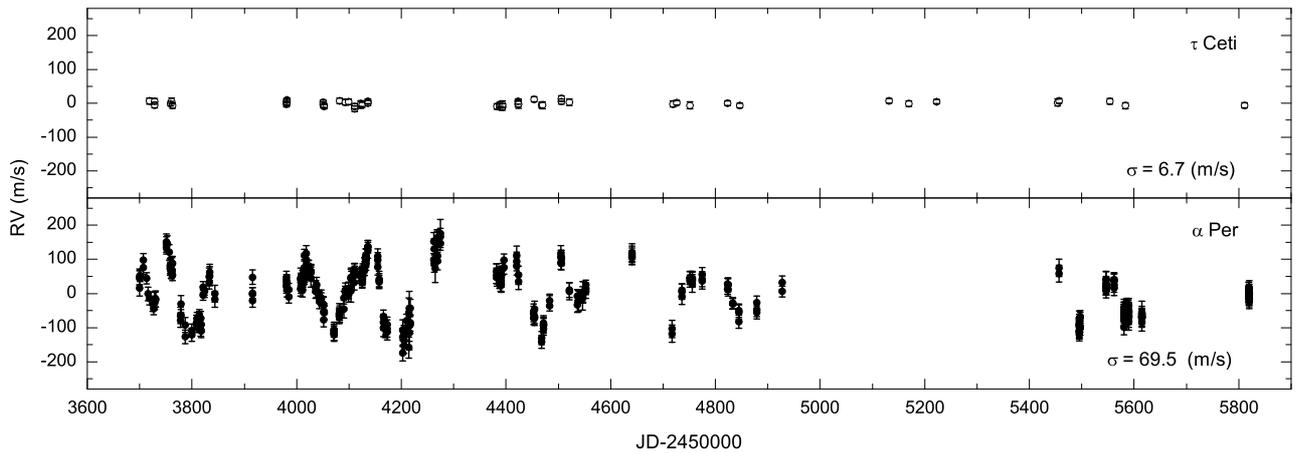}
      \caption{RV measurements of the RV standard star $\tau$ Ceti (\emph{top panel}) and
      $\alpha$ Per from November 2005 to September 2011 (\emph{bottom panel}).
              }
         \label{RV1}
   \end{figure*}

We have acquired 442 spectra of $\alpha$ Per from November 2005 to September 2011 (spanning 133 nights of observations) using the fiber-fed high-resolution (R = 90 000) Bohyunsan Observatory Echelle Spectrograph (BOES; Kim et al. 2007) attached to the 1.8-m telescope at Bohyunsan Optical Astronomy Observatory (BOAO) in Korea. An iodine absorption cell (I$_{2}$) was used to provide the precise RV measurements. Each estimated signal-to-noise ratio (S/N) at the I$_{2}$ wavelength region is about 250 with typical exposure times ranging between 60 and 180 seconds.
The RV measurements of $\alpha$ Per are listed as online data (Table~4).

The extraction of normalized 1--D spectra was carried out using the IRAF (Tody 1986) software. The I$_2$ analysis and precise RV measurements were undertaken using a code called RVI2CELL (Han et al. 2007), which was developed at the Korea Astronomy $\&$ Space Science Institute (KASI).

Figure \ref{RV1} shows RV measurements of $\alpha$ Per and the standard RV star $\tau$ Ceti to demonstrate the long-term stability of the BOES. The figure shows that the RV of $\tau$ Ceti is constant with an rms scatter of 6.7 m s$^{-1}$ over the timespan of our observations.

\begin{figure}
   \centering
   \includegraphics[width=8cm]{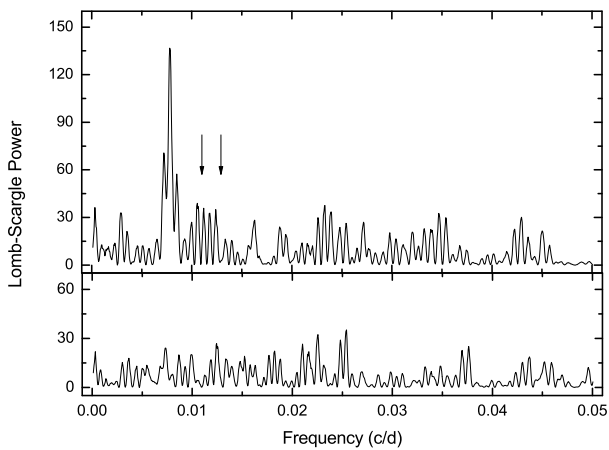}
      \caption{Lomb-Scargle periodogram of the RV measurements of $\alpha$ Per.
      The periodogram shows a significant power at a frequency of 0.0078 c d$^{-1}$ corresponding to a period of 128 days (\emph{top panel}) after subtracting the main frequency variations (\emph{bottom panel}).
      Arrows indicate previously reported periods of 87.8 and 77.7 days.
              }
         \label{power1}
   \end{figure}

We calculated the Lomb-Scargle periodogram of the BOAO RV time series of $\alpha$ Per, which is a useful tool to investigate long-period variations for unequally spaced data (Lomb 1976; Scargle 1982). Figure~\ref{power1} shows a significant power at $f_{1}$ = 0.0078 c\,d$^{-1}$ ($P$ = 128.2 days). We find a false-alarm probability (FAP) of $<$  $10^{-6}$ by a bootstrap randomization process (K{\"u}rster et al. 1999). We could find neither periods of 87.7 days (Hatzes \& Cochran 1995) nor 77.7 days (Butler 1998).

%
%

%

\section{The origin of the RV variations}

Giants and supergiants exhibit pulsations as well as surface activities, resulting in low-amplitude RV variability on different time scales. Short-term (hours to days) RV variations have been known to be the result of stellar pulsations (Hatzes \& Cochran 1998), whereas long-term (hundreds of days) RV variations with a low-amplitude may be caused by three kinds of phenomena:
stellar oscillations, rotational modulations, or planetary companions.
To establish the origin of the low-amplitude and long-periodic RV variations of $\alpha$ Per we examined
the Ca II H \& K lines, the Hipparcos photometry, spectral line bisectors, and fitted a Keplerian orbit to RV data.

\subsection{Ca II H \& K lines}

\begin{figure}
   \centering
   \includegraphics[width=8cm]{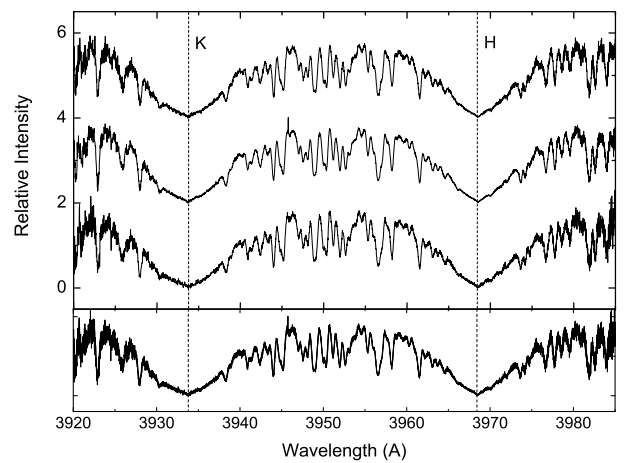}
      \caption{The spectral line of the Ca II H \& K region in $\alpha$ Per.
       It displays three spectra exhibiting positive, zero, and negative RV deviations, respectively (\emph{top panel}), and a combined figure of the three features (\emph{bottom panel}).
      No emission features are seen around the Ca II H \& K central region, and any systematic differences in line profiles are invisible.
        }
        \label{CaII}
   \end{figure}

We used the averaged Ca II H \& K line profiles at the upper, near zero, and lower part of the RV curve to check for any systematic difference in the Ca II H \& K region related to the RV variations.
We selected spectra spanning a long time interval:
the first positive deviation on JD-2453751.125798 (15 January 2006, RV = 144.2 m s$^{-1}$),
near zero deviation on JD-2454535.956330 (10 March 2008, RV = -- 7.0 m s$^{-1}$),
and the last negative deviation on JD-2455495.383283 (25 October 2010, RV = -- 109.3 m s$^{-1}$) of RV curve.
Figure~\ref{CaII} shows that the Ca II H \& K cores show no emissions at the line centers, and there is also no systematic difference among the line profiles at the three extreme parts of the RV curve.
It means that F5 Ib $\alpha$ Per has exhibited no detectable peculiarities in Ca II H \& K lines relevant to RV variations. This agrees with Pasquini et al. (2000), who found that supergiants in their sample of stars show no signatures of Ca II H \& K chromospheric activity.

%
\subsection{Hipparcos photometry}

\begin{figure}
   \centering
   \includegraphics[width=8cm]{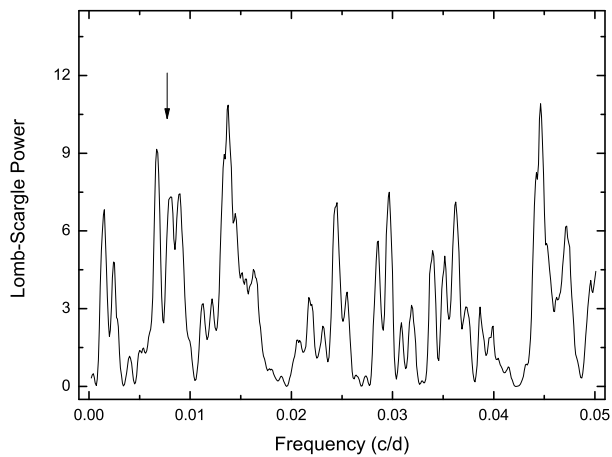}
      \caption{Lomb-Scargle periodogram of the Hipparcos photometric measurements of $\alpha$ Per for three years.
      The arrow marks the location of the period of 128 days.
              }
         \label{Hip}
   \end{figure}

We analyzed the Hipparcos photometry of $\alpha$ Per to search for possible brightness variations relevant to the period of 128 days. For three years, the Hipparcos satellite obtained 87 photometric measurements of $\alpha$ Per.
Figure~\ref{Hip} shows the Lomb-Scargle periodogram of the measurements. There are no significant peaks near the frequency of $f_{1}$ = 0.0078 c d$^{-1}$ corresponding  to the period of 128 days.
The Hipparcos data maintained a photometric stability down to rms of 0.0048 magnitude (0.26\%) during those three years.

\subsection{Line bisector variations}

The analysis of a line shape can help to determine the origin of the observed RV variations (Queloz et al. 2001).
The RV variations by planetary companions should not produce any changes in the spectral line shape, whereas
the surface inhomogeneities or pulsations do. Stellar rotational modulations of surface features or non-radial pulsations can create variable asymmetries in the spectral line profiles.
Thus, the accurately measured variations in the shapes of stellar lines may provide valuable information about surface inhomogeneities, pulsations, or planetary companion. The bisector velocity span (BVS) is defined as the difference between two bisectors at the top and bottom of the profile and is used to quantify the changes in the line profile.

We measured the BVS using the least-squares deconvolution (LSD) technique (Donati et al. 1997; Reiners \& Royer 2004; Glazunova et al. 2008). We used the Vienna Atomic Line Database (VALD; Piskunov et al. 1995) to prepare the list of spectral lines. A total of $\sim$ 3100 lines within the wavelength region of 4500 -- 4900 and 6450 -- 6840 {\AA} were used to construct the LSD profile, which excluded spectral regions around the I$_{2}$ absorption region, hydrogen lines, and regions with strong contamination by terrestrial atmospheric lines.

\begin{figure}
   \centering
   \includegraphics[width=8cm]{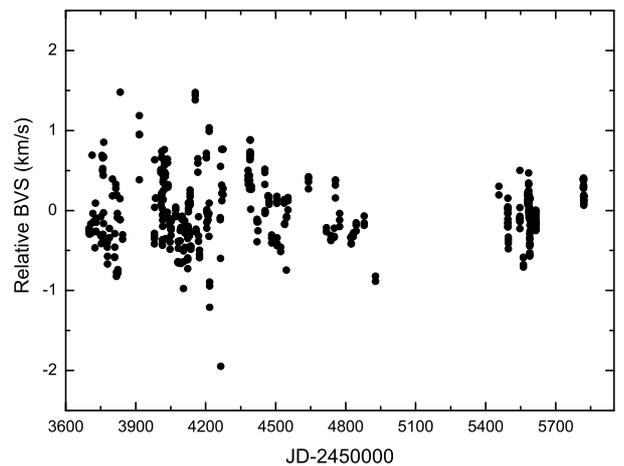}
      \caption{BVS variations of $\alpha$ Per from November 2005 to September 2011.
        }
        \label{LSD}
   \end{figure}

\begin{figure}
   \centering
   \includegraphics[width=8cm]{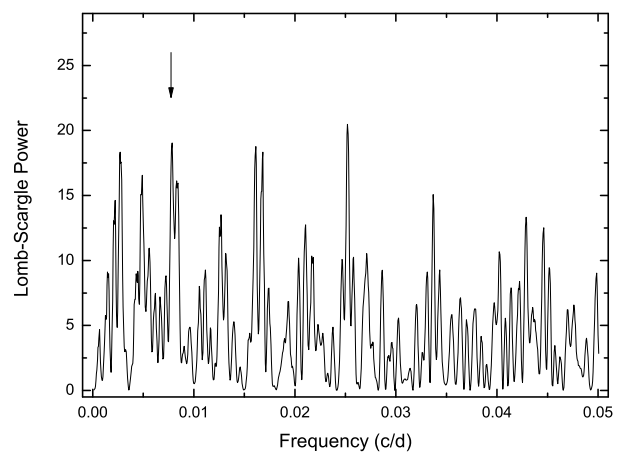}
      \caption{Lomb-Scargle periodogram of BVS variations of $\alpha$ Per.
      The arrow marks the location of the period of 128 days
        }
        \label{LSD_power}
   \end{figure}

We estimated the BVS of the mean profile between two different flux levels of central depth levels, 0.8 and 0.25, as the span points. The BVS as a function of JD is shown in Figure~\ref{LSD}. We calculated the Lomb-Scargle periodogram of the BVS shown in Figure~\ref{LSD_power}. We see a significant peak frequency 0.0078 c d$^{-1}$ (a FAP $\sim$ 0.5\%) relevant to the period of 128 days. There are indeed several significant peaks with higher or comparable power in the periodogram. Although it is difficult to claim with certainty whether the signal near the 128-d period is physically real, it is unlikely that noise would cause this peak. The variations in the BVS might have a higher magnitude compared with the signal at 128 days and hinder it.

%
%

%
\subsection{Keplerian orbital fit}

\begin{figure}
   \centering
   \includegraphics[width=8cm]{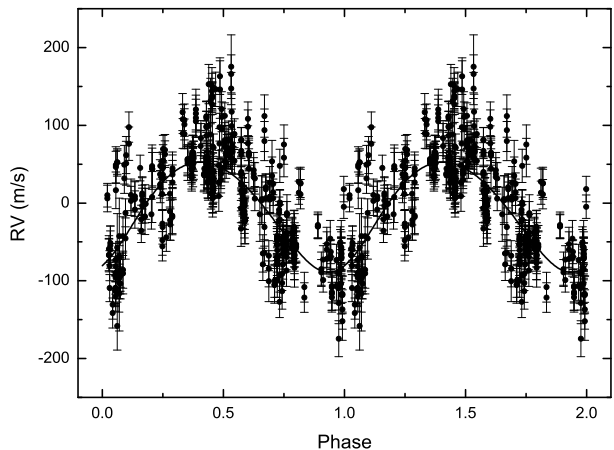}
      \caption{RV measurements of $\alpha$ Per phased to the orbital period of 128 days obtained at BOAO.
      The solid line is the orbital solution that fits the data with an rms of 44.8 m s$^{-1}$.
              }
         \label{phase1}
   \end{figure}

We analyzed our BOAO data using the Lomb-Scagle periodogram and found the best-fitting Keplerian orbit with a $P$ = 128.2 $\pm$ 0.1 days, a semi-amplitude $K$ = 70.8 $\pm$ 1.5 m s$^{-1}$, and an eccentricity $e$ = 0.10 $\pm$ 0.04.
The RV curve phased to the period of 128 days is shown in Figure~\ref{phase1}, and it exhibits clear variations.
We derive the minimum mass of a companion $m$ sin $i$ = 6.6 $\pm$ 0.2 $M_{\mathrm{Jup}}$ at a distance of $a$ = 0.97 AU from $\alpha$ Per, only four times the stellar radius ${R_{\star}}$ = 51 -- 55 $R_{\odot}$ ($\sim$ 0.25 AU).

As can be seen in Figure~\ref{phase1}, the rms of the RV residuals are 44.8 m s$^{-1}$, which is larger than the typical internal error of individual $\alpha$ Per RV accuracy ($\sim$ 20 m s$^{-1}$). We also noticed some systematic pattern in the periodogoram of RV residuals after removing the best fit. Figure~\ref{power1} (bottom panel) shows the Lomb-Scargle periodogram. We found a FAP of $\sim$ $10^{-3}$ at the highest peak of the residual RV measurements, but there are several peaks at this level, which suggests noise.
Irregular periodic RV variations in $\alpha$ Per appear with a 2K amplitude of $\sim$ 150 m s$^{-1}$ in a short period of time. The scatter seems to be caused by variations due to the intrinsic stellar variability or stellar oscillations. All the orbital elements from the BOAO data set are listed in Table~\ref{tab3}.

%
%

%
\begin{table}
\begin{center}
\caption{Best-fitting Keplerian orbit of the hypothetical planet $\alpha$ Per b.}
\label{tab3}
\begin{tabular}{lc}
\hline
\hline
    Parameter                            & Value                           \\

\hline
    Period [days]                        & 128.2  $\pm$ 0.1                \\
    $\it{K}$ [m s$^{-1}$]                & 70.8  $\pm$ 1.5                 \\
    $\it T$$_{\mathrm{periastron}}$ [JD] & 2451513.00 $\pm$ 4.91 (1999.912)  \\
    $\it{e}$                             & 0.10  $\pm$ 0.04                \\
    $\omega$ [deg]                       & 208.37 $\pm$ 12.36              \\
    $f(m)$ [$\it M_{\odot}$]             & (4.6473) $\times$ 10$^{-9}$      \\
    $\sigma$ (O-C) [m s$^{-1}$]          & 44.8                            \\
\hline
    with M = 7.3 [$\it M_{\odot}$]  \\
    $m$ sin $i$ [$\it M_\mathrm{Jup}$]          & 6.6 $\pm$ 0.2                   \\
    $\it{a}$ [AU]                         & 0.97                           \\
\hline

\end{tabular}
\end{center}
\end{table}
%


\section{Discussion}

F-G yellow supergiants are very important for studying the Galactic structure owing to their high luminosity and young age (Klochkova \& Panchuk 1988; Giridhar et al. 1998). Furthermore, all types of supergiants are good science cases in pulsations regarding the Cepheid instability strip. According to the the traditional criteria (Iben \& Tuggle 1975; Fernie 1990), $\alpha$ Per is located outside the blue edge of the Cepheid instability strip
(Hatzes \& Cochran 1995; Bulter 1998).

If supergiants with luminosities and temperatures similar to the Cepheids do not pulsate in the Cepheid instability strip, they may be related to non-variable supergiants (NVSs), which were investigated by Fernie \& Hube (1971).
The presence of the NVSs in the instability strip is quite enigmatic.
s-Cepheids are pulsating stars with low amplitude variations of the light, color, and RV.
They can be considered as objects in the stage between normal Cepheids and the NVSs in the instability strip.
Andrievsky \& Kovtyukh (1996) found no significant differences in the C and O abundances between them and concluded  that these two groups of stars have undergone the first stage of core helium burning and are at the same evolutionary stage. From the photometrist's point of view, $\alpha$ Per can be classified as an NVS near or on the blue border of the Cepheid instability strip.

In this work, we found compelling evidence for a low-amplitude and long-periodic RV variation in the F supergiant $\alpha$ Per. A strongly periodic variability in RV measurements was detected with a period of 128 days.
We will examine possible origins of these RV variations in the next subsections.

\subsection{Pulsations}

The estimated period of the fundamental radial mode pulsations is several days. Therefore, the 128-d period of periodic variations found in $\alpha$ Per cannot be caused by classical $\kappa$-mechanism pulsations operating in Cepheids. On the other hand, we found additional 119-d RV variations on top of the low-amplitude Cepheid variations of 3.97-d period in Polaris (Lee et al. 2008), a Cepheid that lies inside the instability strip. Lee at al. (2008) speculated about the existence of an yet unrecognized mechanism that drives low-amplitude and long-period oscillatory variations of the radius in F-supergiants and in all classical Cepheid stars. Lee et al. (2008) expressed the need for a detailed comparative study of the precise RV for a sample of F-type supergiant stars including Cepheids that can provide additional constraints on the nature of the long-term radius variations detected in Polaris. $\alpha$ Per is the second supergiant star in our survey that shows nearly identical (in terms of RV period and amplitude) long-period quasi-periodic variations and seems to support the assumptions made by Lee et al. (2008) about the common nature of the oscillatory variations of the radius in F-supergiants.

%
   \begin{figure}
   \centering
   \includegraphics[width=8cm]{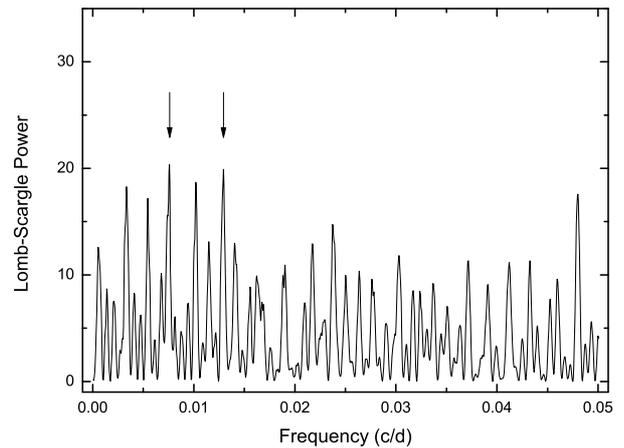}
      \caption{Lomb-Scargle periodogram of the data combined with the McDonald (Hatzes \& Cochran 1995) and Lick observatory (Butler 1998) RV measurements of $\alpha$ Per. Arrows indicate $\sim$ 128 and $\sim$ 77 days.
              }
         \label{H+B}
   \end{figure}
%
%

   \begin{figure}
   \centering
   \includegraphics[width=8cm]{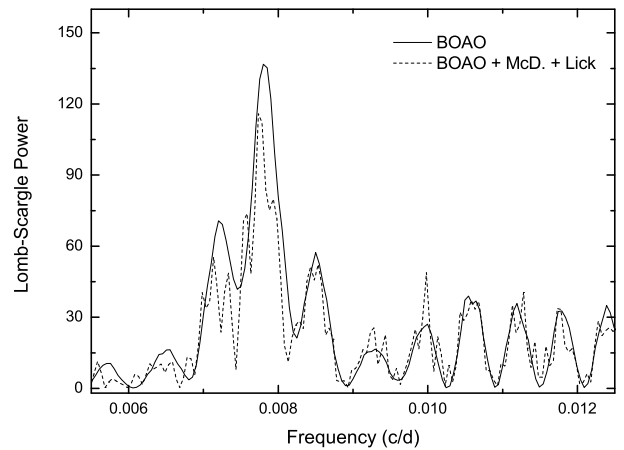}
      \caption{Lomb-Scargle periodograms of the BOAO, McDonald, and Lick observatory RV measurements of $\alpha$ Per. The periodograms show a frequency of 0.0078 c d$^{-1}$ at the BOAO data set (solid line)
      and 0.0077 c d$^{-1}$ at the BOAO + McDonald + Lick observatory data set (dashed line), respectively.
              }
         \label{power2}
   \end{figure}

\subsection{Rotational modulations}


   \begin{figure*}
   \centering
   \includegraphics[width=10cm]{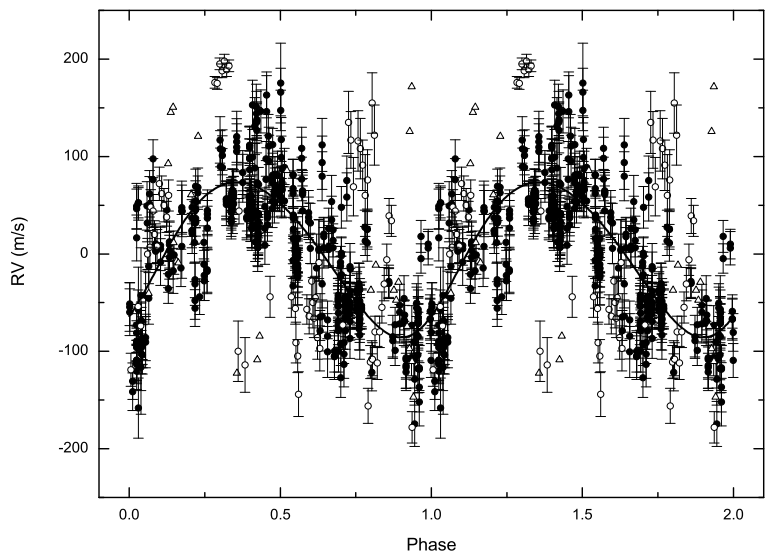}
      \caption{RV measurements of $\alpha$ Per phased to the orbital period of 129.9 days.
      Filled circles, triangles,  and open circles denote the RV measurements of BOAO, McDonald, and Lick observatory, respectively. The solid line is the orbital solution that fits the data with an rms of 58.5 m s$^{-1}$.
              }
         \label{phase2}
   \end{figure*}

Hatzes \& Cochran (1995) and Butler (1998) argued that the cause of the RV variations of $\alpha$ Per is not likely to be the orbital motion, but rather the intrinsic stellar variations such as rotational modulation or non-radial pulsations. The 119-d period with a 2K amplitude of $\sim$ 280 m s$^{-1}$ variations in the residual RV measurements in the F supergiant Polaris could also have been caused by rotational modulation of surface features (Lee et al. 2008), for example.

Supergiants typically have rotational periods of several hundred days (Rao et al. 1993). $\alpha$ Per shows a projected rotational velocity of 17 -- 20 km s$^{-1}$ (see the Table 2) corresponding to a period of 129 -- 139 days (upper limit), very close to the observed period of 128 days. It is also close to the mean projected rotational velocity of 15 km s$^{-1}$ for 16 F supergiants (Danzinger \& Faber 1972). Although the real rotational period is likely to be shorter by the factor of sin $i$ than 129 -- 139 days, it does not eliminate the rotational origin of RV variations. When the inclination of $\alpha$ Per is very close to 90 degrees there is a chance that the observed RV variations are caused by the rotational modulation.

$\alpha$ Per is an object with weak longitudinal magnetic fields of $B_{l}$ = 0.82 $\pm$ 0.37 G (Grunhut et al. 2010) but it exhibits the strongest and most complex Stokes V profile of all 30 supergiant stars observed. Such a low value of magnetic fields  does not mean, however, that the local fields are weak as well. The local magnetic fields can  be stronger and have a complex surface structure. Even a weak magnetic field  in spots is enough to suppress the convection motion and make the velocity field (macroturbulence) inside the spot   lower than the surrounding photosphere. This would alter the line shapes and result in an RV shift. As shown for Polaris by Hatzes \& Cochran (2000) such macroturbulent spots can account for RV variations of up to $\sim$ 100 m s$^{-1}$. Such spots would produce no detectable chromospheric activity indicator (Figure~\ref{CaII}) or photometric variations (Figure~\ref{Hip}). We need a direct confirmation of the existence of the local magnetic fields and the surface inhomogeneities in $\alpha$ Per.

\subsection{Planetary companion}

The important features of the hypothetical planetary RV variations are the strong period, shape, and phase stability of variations. We clearly found 128-d variations. However, the measurements in the previous records by Hatzes \& Cochran (1995) and by Butler (1998) did not show the same variations. Nevertheless, the data combined with previous records do not show a significant power, just two somewhat high peaks exhibited near $\sim$ 128 and $\sim$ 77 days, with a moderate power of $\sim$ 20 (Figure~\ref{H+B}). If we have regular, long-term stable 128-d variations in two data sets, the combination of them should increase the power in the resulting periodogram, which is not the case. We also calculated the Lomb-Scargle periodogram of the whole RV measurements of BOAO, McDonald (Hatzes \& Cochran 1995), and Lick observatory (Butler 1998). Figure~\ref{power2} shows the periodogram of the combined data against that of BOAO. We find a significant power at at $f_{1}$ = 0.0077 c\,d$^{-1}$ ($P$ = 129.9 days) from the whole data set, approximately the same frequency in BOAO data, and an FAP of this periodicity is less than 10$^{-5}$. However, the Lomb-Scargle power decreases, which means that 128-d variations have not been stable on the long-term scale. This is a strong criterion to reject the exoplanet hypothesis. The RV measurements of the whole data set phased to the period of 129.9 days are shown in Figure~\ref{phase2}. It does not conform perfectly with whole RV measurements.

Assuming an exoplanet (least possible) hypothesis, we can estimate the minimum mass for the planetary companion of 6.6 $\it M_{\mathrm{Jup}}$ with an orbital semi-major axis of 0.97 AU and an eccentricity of 0.1.


\begin{acknowledgements}
      Support for MGP was provided by the National Research Foundation of Korea
      to the Center for Galaxy Evolution Research.
      We thank the developers of the Bohyunsan Observatory Echelle Spectrograph (BOES) and
      all staff of the Bohyunsan Optical Astronomy Observatory (BOAO).
      This research made use of the SIMBAD database, operated at CDS, Strasbourg, France.

\end{acknowledgements}
%


\Online
\longtab{4}
{
\begin{longtable}{ccccccccc}
\caption{\label{rv} RV measurements for \object{$\alpha$ Per} during November 2005 and September 2011.}\\
\hline\hline
JD-2 450 000& $\Delta$RV& $\pm \sigma$ & JD-2 450 000& $\Delta$RV& $\pm \sigma$ & JD-2 450 000 & $\Delta$RV& $\pm \sigma$  \\
  day	    & m\,s$^{-1}$ & m\,s$^{-1}$ &     day    & m\,s$^{-1}$ & m\,s$^{-1}$& day	   &   m\,s$^{-1}$ & m\,s$^{-1}$  \\

\hline
\endfirsthead
\caption{continued.}\\
\hline\hline

JD-2 450 000 & $\Delta$RV& $\pm \sigma$ & JD-2 450 000 & $\Delta$RV& $\pm \sigma$ & JD-2 450 000 & $\Delta$RV& $\pm \sigma$  \\
  day	     & m\,s$^{-1}$ & m\,s$^{-1}$ &     day    & m\,s$^{-1}$ & m\,s$^{-1}$ & day      &	 m\,s$^{-1}$ & m\,s$^{-1}$   \\
\hline
\endhead
\hline
\endfoot
3700.264218 &   48.9  &  23.6 &  4104.136653 &       7.7  &     32.7 & 4736.232933 &     8.5 &   20.1  \\
3700.273951 &   46.3  &  20.3 &  4104.144234 &      47.5  &     27.2 & 4736.236603 &     5.2 &   22.2  \\
3700.277782 &   16.4  &  23.5 &  4104.193629 &      43.6  &     21.5 & 4736.240283 &    -8.3 &   23.9  \\
3701.014261 &   52.5  &  15.4 &  4109.897487 &      31.9  &     21.0 & 4751.062019 &    45.1 &   18.0  \\
3701.025690 &   52.2  &  17.5 &  4109.899281 &      54.7  &     20.4 & 4751.065688 &    36.4 &   18.1  \\
3707.207578 &   76.3  &  21.5 &  4109.901168 &      69.4  &     17.7 & 4751.069369 &    46.2 &   19.5  \\
3707.211637 &   97.7  &  19.5 &  4110.905988 &      62.9  &     17.0 & 4756.057215 &    26.5 &   22.4  \\
3713.107931 &   45.1  &  16.5 &  4110.908187 &      68.2  &     20.0 & 4756.060977 &    36.7 &   18.1  \\
3716.077638 &   -1.0  &  19.9 &  4122.962052 &      55.6  &     19.5 & 4756.064658 &    44.6 &   21.1  \\
3719.038270 &  -14.4  &  19.2 &  4122.968614 &      50.7  &     21.6 & 4756.068396 &    40.9 &   20.9  \\
3725.033985 &  -31.0  &  21.7 &  4122.971485 &      53.7  &     18.6 & 4774.375667 &    55.4 &   20.7  \\
3726.937450 &  -43.9  &  18.3 &  4122.974089 &      56.6  &     17.8 & 4774.379336 &    35.6 &   22.8  \\
3728.978831 &  -22.0  &  25.8 &  4125.051986 &      36.2  &     18.1 & 4774.383017 &    41.9 &   22.2  \\
3728.983137 &  -27.7  &  20.1 &  4125.054150 &      61.5  &     16.8 & 4823.007943 &    27.5 &   17.4  \\
3730.134114 &  -17.9  &  23.0 &  4125.056071 &      45.6  &     21.3 & 4823.011832 &    13.4 &   14.9  \\
3730.138929 &  -16.2  &  23.1 &  4125.057992 &      42.9  &     16.5 & 4823.893843 &    25.4 &   20.6  \\
3751.066777 &  151.2  &  23.0 &  4125.059902 &      56.2  &     20.2 & 4823.896331 &    11.4 &   18.9  \\
3751.125798 &  144.2  &  22.6 &  4125.122814 &      48.9  &     15.7 & 4823.898819 &    25.4 &   18.1  \\
3751.130916 &  135.1  &  18.8 &  4125.125395 &      32.8  &     17.3 & 4833.091408 &   -30.0 &   15.8  \\
3753.009406 &  146.4  &  22.1 &  4125.128114 &      53.2  &     20.0 & 4833.093225 &   -28.1 &   16.4  \\
3756.174246 &  120.9  &  22.4 &  4126.042083 &      63.6  &     23.1 & 4833.095042 &   -30.4 &   15.4  \\
3757.963394 &   80.8  &  22.6 &  4126.043912 &      51.7  &     21.0 & 4844.927981 &   -83.0 &   18.9  \\
3757.968301 &   73.1  &  19.3 &  4126.045636 &      57.6  &     21.9 & 4844.930492 &   -50.4 &   19.9  \\
3758.981747 &   70.7  &  17.4 &  4126.047534 &      60.4  &     19.3 & 4844.933003 &   -55.3 &   21.3  \\
3758.990099 &   81.7  &  17.8 &  4129.927865 &      73.4  &     20.2 & 4879.048571 &   -26.6 &   18.4  \\
3760.016120 &   66.8  &  21.9 &  4129.929427 &      83.6  &     22.6 & 4879.051128 &   -55.6 &   19.0  \\
3760.020124 &   81.0  &  24.6 &  4129.930990 &      83.6  &     18.7 & 4879.053686 &   -47.9 &   19.3  \\
3760.958987 &   61.0  &  17.5 &  4132.899083 &      86.4  &     19.0 & 4927.954930 &    31.8 &   20.1  \\
3761.988862 &   65.8  &  18.6 &  4132.902069 &      97.5  &     20.5 & 4927.956377 &     5.6 &   15.5  \\
3761.992560 &   87.3  &  18.1 &  4132.904742 &     110.6  &     19.9 & 5456.333646 &    58.5 &   24.7  \\
3763.033354 &   88.7  &  19.8 &  4132.907346 &      98.8  &     19.2 & 5456.334341 &    75.3 &   25.3  \\
3763.039009 &   54.4  &  16.9 &  4132.910066 &     111.5  &     18.7 & 5495.375111 &   -93.4 &   17.2  \\
3778.065669 &  -64.8  &  24.0 &  4132.912647 &     109.1  &     22.1 & 5495.377206 &   -96.0 &   21.2  \\
3778.070542 &  -78.8  &  20.8 &  4135.908094 &     137.9  &     18.1 & 5495.378734 &  -113.3 &   20.1  \\
3778.922668 &  -30.9  &  24.5 &  4135.909911 &     126.9  &     23.2 & 5495.380250 &  -112.0 &   19.5  \\
3778.926984 &  -63.8  &  16.5 &  4135.911728 &     136.6  &     18.4 & 5495.381767 &  -109.1 &   20.5  \\
3786.907949 &  -92.6  &  21.9 &  4154.916961 &      76.9  &     22.3 & 5495.383283 &  -109.3 &   20.6  \\
3786.913157 & -127.1  &  20.3 &  4154.918906 &      98.3  &     21.5 & 5495.384799 &  -120.3 &   18.8  \\
3800.015511 & -121.5  &  19.1 &  4154.920780 &     108.4  &     21.7 & 5495.386316 &   -89.3 &   20.7  \\
3800.021147 & -108.2  &  19.1 &  4157.909561 &      37.6  &     20.7 & 5495.387832 &   -91.4 &   18.5  \\
3808.968564 &  -89.2  &  16.5 &  4157.911817 &      42.4  &     22.6 & 5495.389348 &   -81.3 &   19.9  \\
3808.974039 &  -85.1  &  22.0 &  4157.913935 &      33.5  &     18.8 & 5495.390968 &   -72.9 &   21.1  \\
3809.938211 &  -72.6  &  17.8 &  4165.943815 &    -100.6  &     26.2 & 5497.395529 &  -102.6 &   20.5  \\
3809.943798 &  -99.0  &  15.5 &  4165.946881 &     -84.5  &     23.2 & 5497.396814 &   -96.3 &   17.8  \\
3813.954326 &  -69.3  &  16.4 &  4165.950157 &     -67.7  &     19.3 & 5497.397751 &   -68.3 &   18.9  \\
3813.959835 &  -62.8  &  15.2 &  4168.932591 &    -104.8  &     24.1 & 5497.398689 &   -70.1 &   15.2  \\
3816.973701 & -106.3  &  18.8 &  4168.934246 &    -100.0  &     21.2 & 5546.880106 &    18.0 &   19.5  \\
3816.980772 &  -73.3  &  20.1 &  4168.935890 &    -105.4  &     21.3 & 5546.881842 &     6.9 &   19.9  \\
3817.941070 &  -90.5  &  18.7 &  4172.924507 &     -91.0  &     21.5 & 5546.883011 &    13.4 &   18.0  \\
3817.947366 & -109.7  &  19.7 &  4172.926243 &    -102.8  &     24.8 & 5546.884180 &    23.1 &   17.0  \\
3820.947399 &   18.0  &  16.3 &  4172.927574 &    -113.5  &     23.1 & 5546.885349 &    41.5 &   22.8  \\
3820.952005 &   -4.7  &  15.1 &  4202.943215 &    -128.6  &     26.5 & 5546.886518 &    28.6 &   21.6  \\
3823.956629 &    5.9  &  17.2 &  4202.947185 &    -174.5  &     23.2 & 5546.887687 &    40.6 &   24.2  \\
3823.961698 &   10.1  &  15.2 &  4202.951293 &    -107.2  &     30.5 & 5546.888856 &    15.0 &   20.0  \\
3832.935945 &   49.2  &  24.9 &  4204.937214 &    -136.9  &     25.2 & 5546.890025 &    28.3 &   18.4  \\
3832.942113 &   31.8  &  19.3 &  4204.939864 &    -152.1  &     24.4 & 5546.891194 &    18.7 &   19.3  \\
3833.950825 &   51.1  &  28.2 &  4204.942445 &    -118.2  &     24.0 & 5561.894825 &    25.4 &   21.9  \\
3833.957480 &   61.7  &  23.8 &  4204.945026 &    -117.0  &     23.2 & 5561.896352 &    15.6 &   21.5  \\
3843.931691 &  -17.4  &  23.1 &  4209.946681 &    -109.4  &     17.5 & 5561.897868 &    38.6 &   20.6  \\
3843.937860 &   -1.1  &  24.6 &  4209.950720 &     -80.6  &     19.4 & 5561.899385 &    36.4 &   20.8  \\
3915.307020 &  -21.3  &  19.1 &  4213.944196 &    -158.2  &     30.8 & 5561.900901 &    39.9 &   20.5  \\
3915.311297 &    0.8  &  15.8 &  4213.948547 &     -57.1  &     43.5 & 5580.913014 &   -72.4 &   20.9  \\
3915.315875 &   -1.9  &  19.3 &  4214.948945 &     -95.4  &     44.8 & 5580.914183 &   -53.4 &   20.9  \\
3915.319858 &   47.8  &  21.3 &  4214.952567 &    -116.2  &     48.3 & 5580.915351 &   -69.1 &   19.7  \\
3980.176625 &   36.9  &  18.9 &  4214.957231 &     -42.6  &     48.8 & 5580.916520 &   -53.3 &   20.5  \\
3980.181799 &   23.9  &  17.7 &  4216.945450 &     -90.5  &     26.3 & 5580.917689 &   -46.7 &   21.7  \\
3980.240219 &   47.9  &  17.1 &  4216.948054 &     -85.8  &     23.8 & 5580.918858 &   -45.0 &   19.8  \\
3980.244617 &   20.9  &  20.6 &  4216.950658 &     -42.5  &     24.9 & 5580.920027 &   -48.2 &   17.9  \\
3981.219074 &   11.7  &  21.3 &  4262.310982 &      99.4  &     18.1 & 5580.921196 &   -42.3 &   25.7  \\
3981.222825 &   31.8  &  21.9 &  4262.312915 &     153.0  &     25.3 & 5580.922365 &   -53.3 &   16.7  \\
3985.309197 &  -10.8  &  17.3 &  4263.309016 &      88.1  &     25.2 & 5581.060109 &   -76.7 &   19.1  \\
3985.312863 &   11.9  &  17.0 &  4263.311204 &     131.3  &     29.3 & 5581.061730 &   -75.7 &   20.2  \\
4007.097216 &   12.4  &  20.2 &  4264.305744 &      73.0  &     40.9 & 5581.063246 &   -99.8 &   21.6  \\
4007.102367 &   44.1  &  30.5 &  4268.307138 &     146.1  &     37.0 & 5581.064762 &   -60.7 &   20.4  \\
4008.283631 &   32.8  &  20.5 &  4268.310032 &     163.2  &     23.5 & 5581.066278 &   -78.8 &   17.6  \\
4008.289400 &   40.4  &  19.6 &  4269.305285 &     110.8  &     19.2 & 5583.991513 &   -78.8 &   19.2  \\
4008.300638 &   41.6  &  15.8 &  4269.306766 &      95.2  &     26.6 & 5583.992681 &   -47.7 &   18.7  \\
4008.304955 &   39.3  &  14.1 &  4274.301293 &     166.1  &     24.4 & 5583.993850 &   -52.1 &   17.9  \\
4008.309176 &   20.7  &  19.8 &  4274.303562 &     147.3  &     20.8 & 5583.995019 &   -43.9 &   18.9  \\
4011.172797 &   26.2  &  17.5 &  4274.305807 &     175.5  &     41.0 & 5583.996200 &   -54.1 &   20.0  \\
4011.176356 &    8.8  &  19.4 &  4382.068443 &      64.9  &     22.2 & 5583.997369 &   -49.4 &   19.0  \\
4011.180045 &   48.5  &  19.1 &  4382.073177 &      65.5  &     21.8 & 5583.998537 &   -65.2 &   19.9  \\
4012.258615 &   41.9  &  13.9 &  4382.077969 &      52.3  &     22.9 & 5583.999706 &   -47.6 &   19.4  \\
4012.262224 &   38.9  &  20.2 &  4382.176494 &      50.0  &     21.5 & 5584.000875 &   -59.3 &   18.4  \\
4012.265720 &   37.0  &  15.0 &  4382.181112 &      45.5  &     24.5 & 5584.002044 &   -43.4 &   21.3  \\
4012.269076 &   29.9  &  17.0 &  4388.088347 &      36.0  &     21.4 & 5585.911355 &   -53.3 &   23.4  \\
4014.353056 &   46.6  &  19.4 &  4388.166801 &      38.0  &     21.6 & 5585.912524 &   -37.8 &   22.9  \\
4014.358959 &   76.1  &  23.7 &  4388.272849 &      44.3  &     20.8 & 5585.913693 &   -53.1 &   22.5  \\
4014.370534 &   54.0  &  16.2 &  4389.279420 &      59.6  &     17.3 & 5585.914862 &   -61.9 &   24.1  \\
4014.380673 &  112.1  &  16.9 &  4389.356812 &      50.5  &     17.7 & 5585.916031 &   -50.6 &   21.6  \\
4015.156985 &   64.7  &  16.8 &  4389.996033 &      36.9  &     18.7 & 5585.917200 &   -42.7 &   20.5  \\
4015.160457 &   61.3  &  17.0 &  4390.047540 &      59.6  &     20.4 & 5585.918368 &   -49.5 &   20.4  \\
4015.163987 &   51.6  &  20.0 &  4390.120449 &      39.2  &     20.3 & 5585.919537 &   -69.1 &   19.4  \\
4015.167333 &   62.7  &  15.9 &  4390.156099 &      26.1  &     19.4 & 5585.920706 &   -86.8 &   20.5  \\
4018.042674 &  117.5  &  22.6 &  4390.199516 &      53.7  &     20.0 & 5585.921875 &   -56.8 &   20.5  \\
4018.048010 &   99.1  &  21.8 &  4390.225096 &      50.8  &     19.7 & 5588.904606 &   -57.6 &   18.2  \\
4018.053774 &   69.9  &  20.0 &  4390.263304 &      28.4  &     22.5 & 5588.906122 &   -57.1 &   18.6  \\
4018.059161 &   58.0  &  19.3 &  4390.285636 &      36.6  &     18.8 & 5588.907638 &   -47.0 &   17.7  \\
4018.337553 &   67.6  &  17.7 &  4390.310292 &      26.0  &     22.2 & 5588.909154 &   -55.8 &   18.2  \\
4018.340979 &   73.6  &  16.6 &  4390.333389 &      46.7  &     19.5 & 5588.910671 &   -40.2 &   19.9  \\
4018.344405 &   54.9  &  17.9 &  4390.356098 &      51.4  &     14.9 & 5588.912430 &   -47.9 &   17.1  \\
4023.111979 &   80.9  &  16.4 &  4393.106831 &      71.3  &     17.5 & 5588.913946 &   -37.0 &   18.8  \\
4023.115690 &   81.1  &  16.0 &  4396.305248 &      76.5  &     19.4 & 5588.915462 &   -58.9 &   17.0  \\
4023.119370 &   67.1  &  18.3 &  4396.309137 &      97.6  &     17.9 & 5588.916978 &   -56.4 &   20.4  \\
4023.122982 &   62.2  &  15.6 &  4420.134262 &     111.9  &     27.2 & 5588.918494 &   -52.2 &   18.6  \\
4027.084998 &   44.2  &  25.5 &  4420.136947 &      79.7  &     25.2 & 5590.082499 &   -62.0 &   20.3  \\
4027.089269 &   63.0  &  14.4 &  4420.139870 &      93.7  &     27.4 & 5590.084362 &   -57.2 &   24.8  \\
4027.092950 &   61.1  &  19.8 &  4424.088364 &      53.6  &     26.9 & 5590.086226 &   -48.3 &   22.9  \\
4027.096804 &   63.1  &  20.5 &  4424.090181 &      35.1  &     24.1 & 5590.088089 &   -83.1 &   25.0  \\
4027.101180 &   64.8  &  20.9 &  4453.078318 &     -52.7  &     21.4 & 5590.089952 &   -67.7 &   23.0  \\
4036.149659 &    6.3  &  20.7 &  4453.081524 &     -59.9  &     22.8 & 5590.091815 &   -66.3 &   21.4  \\
4036.153158 &   20.3  &  16.7 &  4453.084730 &     -71.7  &     19.8 & 5590.093679 &   -36.7 &   23.0  \\
4036.156353 &   20.4  &  17.3 &  4454.893158 &     -71.3  &     22.8 & 5590.095542 &   -71.7 &   26.9  \\
4036.159513 &   17.6  &  16.0 &  4454.895669 &     -45.8  &     23.4 & 5590.097405 &   -54.1 &   22.8  \\
4038.221880 &   26.0  &  21.1 &  4454.898181 &     -67.8  &     20.5 & 5590.099269 &   -53.2 &   23.8  \\
4038.225627 &    7.8  &  21.2 &  4467.937579 &    -141.5  &     19.3 & 5614.912587 &   -59.5 &   37.1  \\
4038.229181 &    4.5  &  16.5 &  4467.939454 &    -130.6  &     18.8 & 5614.914115 &   -84.8 &   25.0  \\
4043.185848 &   -8.6  &  16.3 &  4472.022170 &     -93.3  &     23.5 & 5614.915978 &   -69.2 &   17.9  \\
4043.189115 &  -12.5  &  14.6 &  4472.026533 &     -93.1  &     20.4 & 5614.917841 &   -66.0 &   22.2  \\
4043.192437 &  -23.4  &  20.9 &  4472.030954 &    -103.8  &     19.4 & 5614.919716 &   -65.4 &   19.9  \\
4047.304018 &  -13.0  &  23.6 &  4472.034981 &     -86.4  &     22.0 & 5614.921626 &   -61.4 &   17.5  \\
4047.307501 &  -15.7  &  17.6 &  4483.086998 &     -35.6  &     15.2 & 5614.923489 &   -68.8 &   17.5  \\
4051.088612 &  -30.6  &  21.4 &  4483.089266 &     -20.8  &     14.0 & 5614.925352 &   -71.1 &   19.9  \\
4051.095221 &  -34.5  &  24.6 &  4483.091153 &     -21.0  &     18.3 & 5614.927227 &   -71.4 &   27.3  \\
4051.102073 &  -77.2  &  21.2 &  4504.962102 &     117.0  &     23.9 & 5614.929090 &   -56.9 &   19.9  \\
4052.089730 &  -56.8  &  19.9 &  4505.091755 &      89.6  &     19.1 & 5819.167964 &    -4.7 &   19.8  \\
4052.093607 &  -55.3  &  19.5 &  4505.094093 &     107.9  &     18.6 & 5819.169168 &     2.8 &   19.6  \\
4052.097554 &  -53.0  &  24.2 &  4505.952087 &      88.9  &     19.2 & 5819.170360 &     6.0 &   15.5  \\
4071.069923 & -114.5  &  23.5 &  4505.954796 &     101.2  &     20.0 & 5819.171564 &    13.8 &   18.3  \\
4071.071752 & -116.6  &  21.2 &  4505.957307 &     106.6  &     20.7 & 5819.172757 &    -8.4 &   20.8  \\
4071.073940 & -108.4  &  18.7 &  4520.923815 &       6.3  &     24.2 & 5819.173949 &    10.2 &   22.8  \\
4071.076428 & -106.7  &  21.9 &  4520.926350 &       9.9  &     22.0 & 5819.175164 &   -13.5 &   19.6  \\
4071.078963 & -121.2  &  19.1 &  4535.956330 &      -7.0  &     18.2 & 5819.176414 &    -1.2 &   24.2  \\
4081.276441 &  -59.0  &  17.1 &  4535.958309 &     -33.0  &     22.8 & 5819.177618 &    -4.5 &   20.9  \\
4081.280226 &  -66.7  &  20.4 &  4537.991880 &     -21.7  &     30.5 & 5819.178822 &     3.4 &   17.8  \\
4081.283525 &  -67.0  &  21.5 &  4537.993894 &     -22.7  &     29.7 & 5820.164120 &    15.8 &   22.5  \\
4082.041085 &  -54.5  &  24.9 &  4545.940856 &     -14.5  &     34.7 & 5820.165637 &    -3.7 &   19.1  \\
4082.044233 &  -51.4  &  22.1 &  4546.925839 &       3.6  &     31.3 & 5820.166609 &   -24.2 &   20.6  \\
4082.047833 &  -59.9  &  24.0 &  4546.928350 &       5.0  &     30.4 & 5820.167570 &    -0.5 &   18.4  \\
4090.035309 &  -45.6  &  26.4 &  4552.955671 &      24.7  &     14.0 & 5820.168542 &   -10.7 &   16.3  \\
4090.038908 &  -12.6  &  18.1 &  4552.957581 &       6.2  &     20.0 & 5820.169514 &   -18.7 &   20.4  \\
4092.996622 &    2.8  &  18.7 &  4552.959502 &      13.9  &     17.0 & 5820.170487 &   -15.1 &   18.7  \\
4093.000684 &   10.5  &  20.4 &  4640.287985 &     120.3  &     26.1 & 5820.171418 &     4.2 &   22.4  \\
4098.901678 &    9.6  &  24.1 &  4640.290566 &     105.3  &     20.8 & 5820.172333 &    -0.1 &   19.7  \\
4098.903757 &    3.5  &  19.4 &  4640.293148 &     109.5  &     24.1 & 5820.173247 &    -3.0 &   16.3  \\
4098.905398 &   12.5  &  20.2 &  4640.295763 &     115.0  &     23.7 & 5820.174162 &   -14.4 &   19.6  \\
4098.906847 &   -3.4  &  22.6 &  4717.135611 &    -103.7  &     25.4 &             &         &         \\
4104.131330 &   44.3  &  26.5 &  4717.138721 &    -117.8  &     24.9 &             &         &         \\

\end{longtable}
}

\end{document}